\documentclass[11pt,tightenlines,twoside,onecolumn,nofloats,nobibnotes,nofootinbib,%
superscriptaddress,noshowpacs,centertags]{revtex4-2}
\usepackage{ljm}
\usepackage{xcolor}
\usepackage{amsmath}
\usepackage{amssymb}
\usepackage{url}
\usepackage{hyperref}
\usepackage[cp1251]{inputenc}
\usepackage[russian]{babel}

\begin{document}
    \titlerunning{Numerical Modeling of Galactic Cosmic Ray Modulation in the Heliosphere}
    \authorrunning{Shestakov \& Izmodenov} 

    \title{Numerical Modeling of Galactic Cosmic Ray Modulation in the Heliosphere}

    \author{\firstname{D. A.}~\surname{Shestakov}}
    \email[E-mail: ]{dashestakov_1@edu.hse.ru}
    \affiliation{National Research University Higher School of Economics (HSE), Moscow, Russia}

    \author{\firstname{V. V.}~\surname{Izmodenov}}
    \email[E-mail: ]{izmod@cosmos.ru}
    \affiliation{Space Research Institute (IKI) of RAS, Moscow, Russia}

\received{August 20, 2025;  revised XXX; accepted XXX}

\begin{abstract}
Numerical modeling of galactic cosmic rays (GCRs) penetration through the heliosphere to the vicinity of the Sun is considered. Galactic cosmic rays are charged particles with energies exceeding 10 MeV/nucl., originating from far beyond the boundaries of our Solar System. As they penetrate through the heliosphere - the region of space filled by the solar wind - they interact strongly with the interplanetary magnetic field. In this paper, we present numerical approaches to solving the so-called Parker transport equation for the isotropic velocity distribution function of GCRs. This equation includes a convective term, anisotropic diffusion, adiabatic cooling, and drifts. Additionally, the diffusion coefficient is spatially and energy-dependent, varying by several orders of magnitude. Our numerical approaches are based on the finite-difference method (Crank-Nicolson scheme) and the stochastic differential equations (SDE) method. The numerical methods were validated against a known analytical solution under simplified conditions.
For the general problem formulation, which involves anisotropic diffusion and the Parker spiral interplanetary magnetic field configuration,  we used  the most efficient and flexible SDE method and compared the numerical results with the data from the works of   Kota \& Jokipii \cite{Jokipii-1983} and Burger \cite{Burger-2012}. Special attention was devoted to incorporating drift along the heliospheric current sheet in the model.
\end{abstract}

\maketitle

\section{INTRODUCTION}
Galactic cosmic rays (GCRs) are charged particles with energies above 10 MeV/nucleon, originating from outside the boundaries of our Solar System. The predominant component of these particles is protons that have been accelerated by shock waves within the Milky Way galaxy. Before reaching Earth and being measured, GCRs pass through the heliosphere and the heliospheric boundary region, where the solar wind interacts with the local interstellar medium.

The gasdynamic structure of the global heliosphere was first proposed qualitatively by Baranov, Krasnobayev, and Kulikovsky \cite{Baranov-1970}. This structure, shown in Figure 1, includes: 1) the heliospheric termination shock (TS), where the solar wind is decelerated from supersonic to subsonic velocities; 2) the tangential discontinuity (called the heliopause), which separates the solar wind from the interstellar plasma; and 3) the bow shock in the interstellar medium. The region between the termination shock and the heliopause is often referred to as the inner shock layer. Modern models of the heliospheric boundaries are 3D kinetic-MHD models that account for the multi-component nature of both the interstellar and solar winds, interplanetary and interstellar magnetic fields, and latitudinal and temporal variations in the solar wind. A state-of-the-art model has been developed by our group [see, for example, \cite{Izmodenov-2015}, \cite{Izmodenov-2020}].

The ultimate goal of our work is to develop a numerical model that allows us to explore the so-called modulation of the GCRs through the global heliosphere. Modulation means the change in GCRs intensity as they propagate through the heliosphere. This problem is highly complex because, as will be shown later, it requires solving the Parker transport equation, which includes a convective term, anisotropic diffusion, adiabatic cooling, and drifts. The coefficients in this equation depend on the plasma and magnetic field distributions and vary significantly in physical and velocity space.

Despite the large number of papers studying GCRs modulation in the heliosphere, most of these studies are restricted to modulation in the supersonic solar wind near the Sun. Only a few explore modulation at the heliospheric boundaries. Izmodenov \cite{Izmodenov-1997} calculated GCRs modulation using a 2D kinetic-gasdynamic model of the global heliosphere by Baranov and Malama \cite{Baranov-1993}. However, since the magnetic field was not included in the global model, an oversimplified approach for the diffusion coefficient was adopted. The most advanced modern models of GCRs modulation, including the heliospheric boundaries, were developed by Florinski et al. \cite{Florinski-2009, Florinski-2011}. However, these models used plasma and magnetic field parameters from alternative (and sometimes oversimplified) global heliosphere models.

In this paper, we describe numerical methods developed for solving the Parker transport equation. For simplified formulations of the problem, we examine the application of the finite-difference approach (Crank-Nicolson scheme) and the stochastic differential equations (SDE) method. The results obtained from both numerical methods were cross-validated against each other and benchmarked against the analytical solution.  For the most general case - incorporating three-dimensional geometry, anisotropic diffusion, and drift effects - we present the implementation of the SDE method and compare the results with previous studies conducted by Kota \& Jokipii \cite{Jokipii-1983} and Burger \cite{Burger-2012}.

\begin{figure}[!h]
\centering
\begin{minipage}{0.49\linewidth}
    \includegraphics[width=\linewidth]{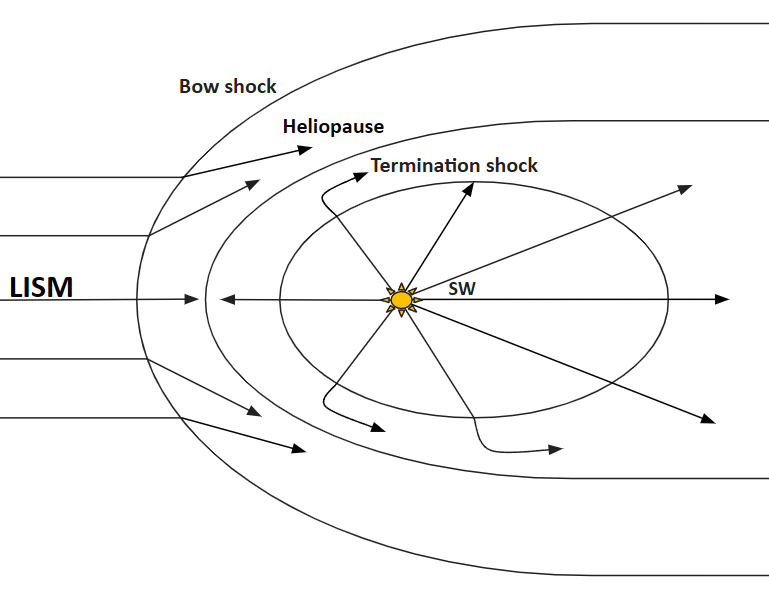}
\end{minipage}%
\hfill
\begin{minipage}{0.49\linewidth}
    \raggedright
    \caption{Schematic representation of the heliospheric shock layer formed by the interaction of solar wind (SW) with the local interstellar medium (LISM): the termination shock (TS), the heliopause, and the bow shock.}
    \label{fig:mpr}
\end{minipage}
\end{figure}

\section{PARKER TRANSPORT EQUATION}
The dynamics of high-energy charged particles in interplanetary space has been thoroughly examined in~\cite{Toptygin-1983}. For GCR propagation through the heliosphere, the dominant effects arise from interactions with interplanetary electromagnetic fields, while gravitational fields, Coulomb scattering, and nuclear interactions can be neglected. Additionally, the GCR energy density is negligible compared to the interplanetary magnetic field (IMF) energy density, allowing the IMF to be treated as a fixed background field. The IMF is <<frozen>> into the solar wind (SW) plasma and contains small-scale random irregularities (comparable to or smaller than the GCR gyroradius), which induce rapid stochastic variations in both the spatial coordinates and energy of the particles.

The transport of GCRs is described by the pitch-angle-averaged distribution function $f(\mathbf{r},p,t)$, governed by the Parker transport equation \cite{Parker-1965}:
\begin{equation}
\frac{\partial f}{\partial t} + (\mathbf{u}+\mathbf{V_{dr}}) \cdot \nabla f - \nabla \cdot(\boldsymbol{\hat{\kappa}}\cdot \nabla f) - \frac{1}{3} (\nabla \cdot \mathbf{u}) p\frac{\partial f}{\partial p} = 0,
\end{equation}
where $\mathbf{u}$ is the SW velocity, $\mathbf{V_{dr}}$ represents the drift velocity of particles for an isotropic velocity-space distribution function, and $\boldsymbol{\hat{\kappa}} = \begin{pmatrix}
     \kappa_{\parallel} & 0 & 0\\
    0 & \kappa_{\perp} & 0\\
    0 & 0 & \kappa_{\perp}
    \end{pmatrix}$ denotes the GCR spatial diffusion tensor in the magnetic field-aligned coordinate system, with $\kappa_{\parallel}$ being the parallel diffusion coefficient (along the magnetic field direction) and $\kappa_{\perp}$ the perpendicular diffusion coefficient.

Thus, the IMF and SW plasma generate the following modulation processes: spatial diffusion due to particle scattering on random small-scale magnetic field irregularities; energy changes including adiabatic energy losses from the expanding SW plasma flow; particle drift caused by IMF gradient and curvature; convective transport of particles by the solar wind plasma.

\section{NUMERICAL METHODS}
This section describes the implementation of both the grid-based finite-difference scheme and the SDE method for a steady-state spherically symmetric case. The analysis incorporates the following simplifying assumptions:
\begin{itemize}
\item The heliosphere is modeled as a sphere with radius $R = 90\ \text{AU}$. The effects of both TS and heliopause are neglected.
\item The SW velocity is assumed constant in magnitude and purely radial: $\mathbf{u} = u \mathbf{e_r}$, where $u = 4\times 10^7$ $\text{cm/s}$.
\item Particle drift $\mathbf{V_{dr}}$ in the inhomogeneous IMF has no radial component.
\item The diffusion tensor $\boldsymbol{\hat{\kappa}}$ is isotropic.
\item The diffusion coefficient follows the form:
\begin{equation*}
\kappa (r, p) = \kappa_0 \left(\frac{p}{p_0}\right)^a \left(\frac{r}{r_0}\right)^b,
\end{equation*} where $\kappa_0 - const$, $\frac{p}{p_0}$ is the particle momentum in units of $\text{GeV/c}$, and $\frac{r}{r_0}$ denotes the radial distance in astronomical units ($\text{AU}$) in the heliocentric coordinate system.
\end{itemize}
Under these conditions, Parker's transport equation (1), governing the cosmic ray distribution function $f(r,p)$, can be expressed as follows:
\begin{equation}
\left (u - {\kappa(r,p)(2+b) \over r}\right)\frac{\partial f}{\partial r}  - {\kappa(r,p)}{\partial ^{2}f \over \partial r ^{2}}  - \frac{2 u}{3r} p\frac{\partial f}{\partial p} = 0
\end{equation}
The considered boundary conditions are:
\begin{equation}
     f(R, p) = (p/p_0)^{-\gamma}, \quad \quad \left.\frac{\partial f(r, p)}{\partial r}\right\vert_{r = r_{\odot}} = 0, \quad \quad f(r, p = p_{\text{max}}) = (p_{\text{max}}/p_0)^{-\gamma}
\end{equation}
At a distance of $R = 90\ \text{AU}$ cosmic rays are assumed to have an unmodulated (in the LISM) power-law energy spectrum $j(r,p) \propto p^{-(\gamma-2)} = 4\pi p^2 f(r,p)$, where the spectral index $\gamma$ typically ranges between  4.5 and 4.7. Near the solar radius at $r_{\odot} = 0.005\ \text{AU}$, a zero particle flux condition is imposed. Additionally, it is assumed that particles with energies up to $p_{\text{max}} = 50\ \text{GeV/c}$ are practically unaffected by heliospheric modulation due to their ultrarelativistic velocities.

\subsection{Grid-based finite-difference scheme (Crank-Nicolson)}
Implementation of the Crank-Nicolson scheme begins with a coordinate transformation. We introduce an effective time variable $t = -\ln{p}$. Since particles lose energy, this transformation ensures that $t$ increases monotonically. In this new variable, equation (2) takes a form analogous to the heat conduction equation:
\begin{equation}
    \frac{\partial f}{\partial t} = A(r,t) {\partial ^{2}f \over \partial r ^{2}} + B(r, t) \frac{\partial f}{\partial r},
\end{equation}
where \quad $A(r,t) = \frac{3r}{2u}\kappa(r,t)$; \quad $B(r,t) = \frac{3}{2}\left(\frac{\kappa(r,t) (2+b)}{u}-r\right)$.

For parabolic equations, particularly the heat conduction equation, the implicit Crank-Nicolson finite-difference scheme \cite{Crank-1947} is typically employed. In this framework, the following derivative approximations must be used:
\begin{equation}
     \left.\frac{\partial ^{2}f}{\partial r ^{2}}\right \vert_{r = r_j} \approx \frac{1}{2} \left(\frac{f_{j+1}^i - 2 f_j^i + f_{j-1}^i}{\delta r^2} + \frac{f_{j+1}^{i+1} - 2 f_j^{i+1} + f_{j-1}^{i+1}}{\delta r^2}\right) ,
\end{equation}
\begin{equation*}
    \left.\frac{\partial f}{\partial r} \right\vert_{r = r_j} \approx \frac{1}{4} \left(\frac{f_{j+1}^i - f_{j-1}^i}{\delta r}+ \frac{f_{j+1}^{i+1} - f_{j-1}^{i+1}}{\delta r}\right), \quad \left.\frac{\partial f}{\partial t}\right \vert_{t = t_i} \approx \frac{f_j^{i+1} - f_j^{i}}{\delta t},
\end{equation*}
where index $j$ corresponds to the radial distance $r$ grid, while index $i$ represents the grid in the effective time variable $t$.
Substituting expressions (5) into equation (4) yields:
\begin{multline}
     \left(\frac{\alpha_j^i}{2} + \beta_j^i\right) f_{j+1}^{i+1} - (1+\alpha_j^i) f_j^{i+1} + \left(\frac{\alpha_j^i}{2} - \beta_j^i\right) f_{j-1}^{i+1} = \\ = -\left(\frac{\alpha_j^i}{2} + \beta_j^i\right) f_{j+1}^{i}  + (\alpha_j^i - 1) f_j^{i} - \left(\frac{\alpha_j^i}{2} - \beta_j^i\right) f_{j-1}^{i}
\end{multline}
where \quad $\alpha_j^i = \frac{A_j^{i} \delta t}{\delta r^2}$, \quad $\beta_j^i = \frac{B_j^{i} \delta t}{4 \delta r}$, \quad $j =  1, 2,  ..., N-2;$ \quad $i = 0, 1, 2, ...,  M-2$.
\\ \\
The values $f_{j-1}^{i+1}$, $f_{j}^{i+1}$, and $f_{j+1}^{i+1}$ are computed by solving the tridiagonal system of linear equations (6) using the tridiagonal matrix algorithm~\cite{Gelfand-1952}.

\subsection{SDE method}
A general formulation of the system of stochastic differential equations (SDE), mathematically equivalent to the Parker equation (1) in backward-in-time representation, is discussed in detail in~\cite{Zhang-1999}. In this formulation (steady-state, spherical symmetry), the system of SDE takes the following form:
\begin{equation}
    d\mathbf{X} = \left(\nabla \cdot \boldsymbol{\hat{\kappa}} - \mathbf{u}\right)ds + \boldsymbol{\alpha} \cdot dW(s), \quad \quad \quad  dp = p\frac{2 u}{3r}  ds,
\end{equation}
\begin{equation*}
    \nabla \cdot \boldsymbol{\hat{\kappa}} = \frac{\kappa(r,p)(2+b)}{r} \mathbf{e}_r, \quad \quad \mathbf{u} = u \mathbf{e}_r, \quad \quad \boldsymbol{\alpha} = \sqrt{2 \kappa(r,p)}\, \mathbf{e}_r,
\end{equation*}
where $s$ represents the time variable in the backward-in-time formulation of the stochastic process; $\mathbf{e}_r$ denotes the unit vector in the radial direction of the heliocentric coordinate system; and the Wiener process increment is defined as $dW(s) = \sqrt{ds}\, N(0,1)$ and represents a normally distributed random variable with zero mean and variance $ds$ (the given time step).

The system of SDE (7) describes the backward-in-time trajectory of so-called pseudo-particles along with the corresponding change in particle momentum magnitude during their passage through the heliosphere. The evolution of coordinates $\mathbf{X}$ is determined by both the regular particle motion (the first term in (7)) and the stochastic component (the second term), which arises due to particle scattering on small-scale irregularities of the IMF. The stochastic component is mathematically represented by a Wiener process, which provides a description of Brownian motion dynamics.

To numerically solve the original problem for the cosmic-ray distribution function, it is necessary to simulate a large ensemble (in our computations, 10000 are used) of backward-in-time trajectories of pseudo-particles. To determine the value of the distribution function at a given point in phase space $(x, y, z, p)$, this point must be used as the initial condition for each pseudo-particle. The system is integrated until the pseudo-particle reaches the heliosphere's outer boundary at $R = 90\ \text{AU}$. Upon reaching the heliospheric boundary, the terminal momentum magnitude $p_{\text{end}}$ is registered. Additionally, the inner boundary at radius $r_{\odot}$ is treated as a reflecting wall to satisfy the zero particle flux boundary condition. 

The numerical solution to the transport equation (2) at the specified phase-space point $(x, y, z, p)$ is evaluated through the following expression:
\begin{equation}
    f(x, y, z, p) = \frac{1}{N} \sum_{k=1}^{N} f_{\text{LISM}}(p_{\text{end},k}),
\end{equation}
where $p_{\text{end},k}$ denotes the momentum magnitude of the $k$-th pseudo-particle at the termination point of its trajectory, $f_{\text{LISM}}(p) \sim p^{-{\gamma}}$ represents the cosmic-ray distribution function in the LISM, while $N$ denotes the total number of simulated trajectories (pseudo-particles) in the numerical computation.

Since the value of $f_{\text{LISM}}$ as well as the problem parameters are time-independent, the solution (8) for the distribution function $f(x, y, z, p)$ is steady-state. In this case, the time variable $s$ serves solely as an internal parameter describing the pseudo-particle trajectory evolution.

An essential requirement is that the SDE integration must be performed in Euclidean space (see, e.g.,~\cite{Gardiner-1983}). Consequently, the spatial coordinates $\mathbf{X}$ are defined in a Cartesian coordinate system. The integration is performed using an explicit numerical scheme: 
\begin{align*}
    x_{i+1} &= x_i + \left( \nabla \cdot \boldsymbol{\hat{\kappa}} - \mathbf{u} \right)^i_x\, ds + \boldsymbol{\alpha}_x^i\, dW, \\
    y_{i+1} &= y_i + \left( \nabla \cdot \boldsymbol{\hat{\kappa}} - \mathbf{u} \right)^i_y\, ds + \boldsymbol{\alpha}_y^i\, dW, \quad \quad \quad  p_{i+1} = p_i + p_i \left( \frac{2u}{3r} \right)^i ds, \\
    z_{i+1} &= z_i + \left( \nabla \cdot \boldsymbol{\hat{\kappa}} - \mathbf{u} \right)^i_z\, ds + \boldsymbol{\alpha}_z^i\, dW, \\
\end{align*}
where $x_i = x(s_i)$, $y_i = y(s_i)$, $z_i = z(s_i)$, and $p_i = p(s_i)$ represent the pseudo-particle's spatial coordinates and momentum magnitude  at time $s_i$; the terms $(\nabla \cdot \boldsymbol{\hat{\kappa}} - \mathbf{u})^i_{x,y,z}$ and $\boldsymbol{\alpha}^i_{x, y, z}$ correspond to the respective $x$, $y$, $z$ components of these vector quantities at time $s_i$.

\section{VERIFICATION \& RESULTS}
A steady-state spherically symmetric formulation of the problem with a constant diffusion coefficient $\kappa(r,p) = \kappa_0 - const$ was thoroughly examined in~\cite{Toptygin-1983}. Such simplifications allow for an analytical solution, which will be used for the verification of numerical schemes.

To visualize the effect of adiabatic cooling, the boundary condition at $R = 90\ \text{AU}$ was modified as follows:
\begin{equation}
f(R,p) =\begin{cases} (p/p_0)^{-\gamma}, & p \geq p_0 \\
                     0, &  p < p_0
       \end{cases}
\end{equation}
Therefore, the analytical solution to Parker's equation (2) takes the form:
\begin{equation}
f(r,p) =
\begin{cases} 
\dfrac{F(2\gamma/3, 2, ur/\kappa_0)}{F(2\gamma/3, 2, uR/\kappa_0)} \left(\dfrac{p}{p_0}\right)^{-\gamma}, & p \geq p_0 \\[15pt]
\dfrac{3}{2} \sum\limits_{n=0}^{\infty}\dfrac{F(b_n, 2, ur/\kappa_0)}{ (\gamma - 3 b_n / 2)F'(b_n, 2, uR/\kappa_0)}  \left(\dfrac{p_0}{p}\right)^{3 b_n / 2}, & p < p_0
\end{cases}
\end{equation}
where $F(\alpha,\beta,z)$ is the degenerate hypergeometric function, $b_n$  are the roots of the equation $F(b_n, 2, uR/\kappa) = 0$, and $F'(...)$ - denotes the derivative with respect to the first argument.

Figure 2 demonstrates excellent agreement between the numerical solution obtained using the Crank-Nicolson scheme (for the case where $a = 0$ and $b = 0$) and the analytical solution (10). Moreover, the plot distinctly reveals the effect of adiabatic cooling: near Earth's orbit, the distribution function becomes non-zero for particles with momentum magnitude $p < p_0 = 1\ \text{GeV/c}$.

\begin{figure}[!h]
\centering
\includegraphics[width=0.55\linewidth]{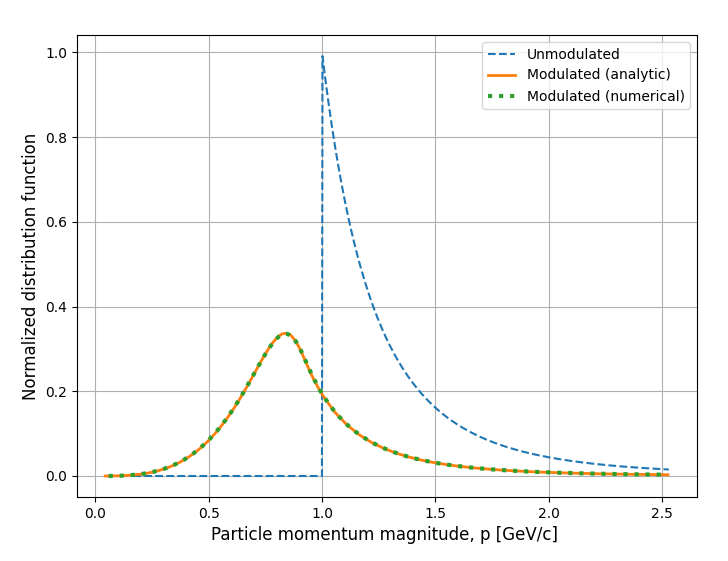}
\caption{Comparison of analytical (110-term series) and Crank-Nicolson numerical solutions at 1 AU (labeled "Modulated"). The unmodulated outer boundary condition (9) is imposed. The diffusion coefficient is constant: $\kappa_0 =  4.5 \times 10^{22}$ $\text{cm}^2/\text{s}$.}
\label{fig:mpr}
\end{figure}

Figure 3 shows the kinetic energy dependence of the GCR energy spectra $j(r,p) \propto p^2f(r,p)$, calculated for an energy-independent diffusion coefficient ($a = 0$). The continuous boundary condition (3) was implemented at the outer boundary ($R = 90\ \text{AU}$). The plot demonstrates that the shape of the GCR energy spectrum remains invariant as particles propagate through the heliosphere when compared to the spectrum in the LISM. Figure 4 illustrates that this invariance is a result of the energy-independent diffusion coefficient. Furthermore, for the case of $a = 2$ shown in Figure 4, the problem was solved using both the Crank-Nicolson scheme (labeled "C-N scheme") and the SDE method, thereby providing cross-verification of numerical methods.

\begin{figure}[!h]
\centering
\begin{minipage}{0.49\textwidth}
\centering
\includegraphics[width=\linewidth]{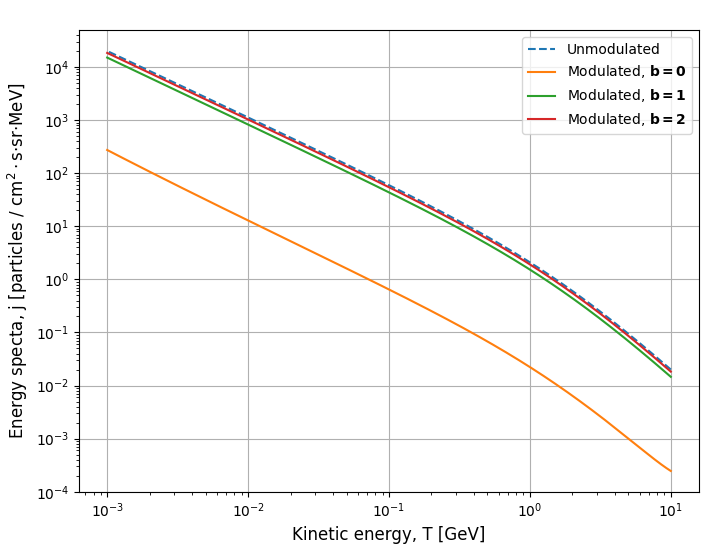}
\caption{GCR energy spectra at 1 AU (modulated) and 90 AU (unmodulated) for the case of diffusion coefficient: $\kappa(r) = \kappa_0 \left(\frac{r}{r_0}\right)^b$, $\kappa_0 = 1.5 \times 10^{22}$ $\text{cm}^2/\text{s}$}
\label{fig:mpr_a}
\end{minipage}
\hfill
\begin{minipage}{0.49\textwidth}
\centering
\includegraphics[width=\linewidth]{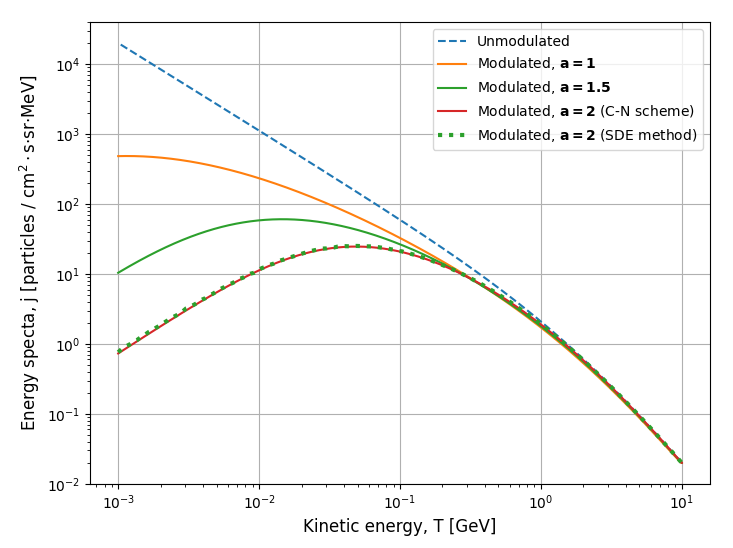}
\caption{Same as Fig. 3 but for the case of diffusion coefficient: $\kappa(r,p) = \kappa_0 \left(\frac{r}{r_0}\right) \left(\frac{p}{p_0}\right)^a$, $\kappa_0 = 1.5 \times 10^{22}$ $\text{cm}^2/\text{s}$}
\label{fig:mpr_b}
\end{minipage}
\end{figure}

For more complex problems, finite-difference methods become computationally expensive, particularly due to their high memory requirements. Furthermore, features in the drift term (discussed in the following section) can compromise numerical stability. Therefore, for general problem formulations and studies of GCR modulation beyond the supersonic SW region, we will implement the SDE method. This method is less sensitive to stability constraints and significantly reduces memory demands. The primary limitation of the SDE method is that it provides solutions only at discrete points in phase space $(x, y, z, p)$, rather than across the entire computational domain, as is typical of finite-difference schemes.

\section{THREE-DIMENSIONAL MODEL}
This section presents the implementation of the SDE method for the most general case, incorporating three-dimensional spatial geometry, anisotropic diffusion, and effects of particle drift in the inhomogeneous IMF. The results are compared with prior studies \cite{Jokipii-1983}, \cite{Burger-2012}, which examined GCR modulation with both flat and wavy configurations of the heliospheric current sheet.

In the simplest model, the IMF is described by the Parker field \cite{Parker-1958}, which can be expressed in heliocentric coordinates through the radial distance $r$, zenith angle $\theta$, and azimuthal angle $\phi$ as follows:
\begin{equation}
\mathbf{B} = A_{\text{pol}} B_0\left(\frac{r_0}{r}\right)^2 (\mathbf{e_r} - \frac{r \Omega_{\odot} \sin{\theta}}{u} \mathbf{e_{\phi}}) \left[1-2 H \left(\theta -\theta_{cs}\right)\right],
\end{equation}
where $A_{\text{pol}} = \pm 1$ is the magnetic field polarity; $B_0 = 35\ \mu \text{G}$ represents the magnetic field magnitude at Earth's orbit: $B_e = \sqrt{2}B_0 \approx 50\ \mu\text{G}$; $\Omega_{\odot} \approx 3 \times 10^{-6}\ \text{s}^{-1}$ denotes the solar rotation angular velocity; $H(\theta - \theta_{\text{cs}})$ is the Heaviside step function, $\theta_{\text{cs}} = \theta_{\text{cs}}(r,\phi)$ defines the heliospheric current sheet surface angle.

The drift velocity, caused by IMF gradient and curvature, in the Parker equation must be calculated under the assumption of an isotropic cosmic-ray distribution function in velocity space. A detailed derivation of this expression is given by Burger et al.~\cite{Burger-1985}. The resulting formula takes the form:
\begin{equation}
    \mathbf{V_{dr}} = \frac{v P}{3} \nabla \times \left( \frac{\mathbf{B}}{B^2} \right), \quad \quad P \equiv \frac{pc}{eZ},
\end{equation}
where $v$ is the particle velocity and $P$ is the magnetic rigidity. The drift velocity (12) calculated from field (11) becomes singular at $\theta_{cs}$ due to the $\delta$-function from differentiating the Heaviside step function. As shown in~\cite{Burger-1985}, these divergences are non-physical artifacts of classical drift theory.

In our calculations for drift velocity, we replaced the abrupt sign reversal $\left[1 - 2H(\theta - \theta_{cs})\right]$ in the magnetic field expression (11) with a smoother $\tanh[k(\theta_{cs} - \theta)]$ function which recovers the original formulation in the limit $k \to \infty$. This finite smoothing parameter $k = k_0 \cdot (p_0/p)$ eliminates the current sheet singularity by effectively introducing a finite thickness, where the choice $k \
\sim 1/R_g$ reflects the particle gyroradius-dependent influence region of the drift singularity near the sheet, consistent with Burger et al.~\cite{Burger-1985}. The $k_0$ value is calibrated to match drift velocities for the case of $\theta_{cs} = \pi/2$ from the Burger et al.~\cite{Burger-1989} reference model. Figure 5 demonstrates close agreement between our alternative approach and the Burger et al.~\cite{Burger-1989} results near the current sheet.

\begin{figure}[!h]
\centering
\includegraphics[width=0.55\linewidth]{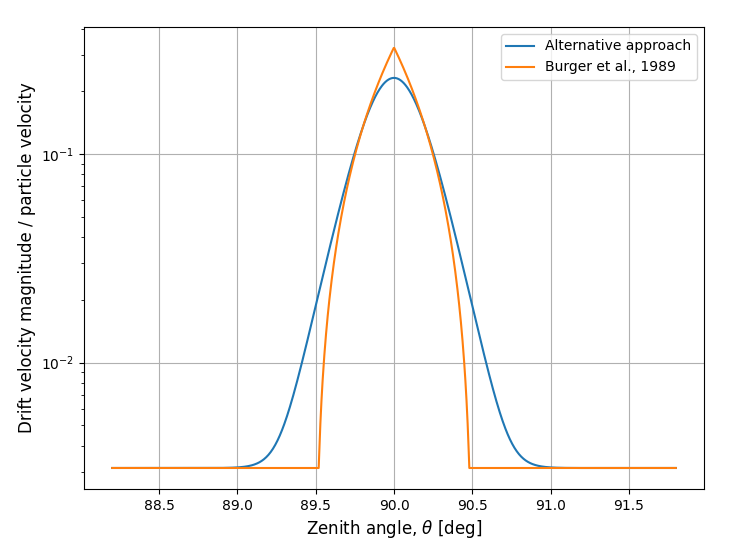}
\caption{Comparison of normalized drift velocity magnitude ($|\mathbf{V_{dr}}|/v$) between the proposed alternative approach and the Burger et al.~\cite{Burger-1989} flat current sheet model ($\theta_{cs} = \pi/2$), shown as a function of zenith angle at $r = 1\ \text{AU}$ for $p = 1\ \text{GeV/c}$ particles.}
\label{fig:mpr}
\end{figure}

The wavy configuration of the heliospheric current sheet is modeled under the assumptions of constant magnitude and purely radial SW velocity using the analytical formulation derived by Kota \& Jokipii \cite{Jokipii-1983}, which specifies  the sheet's position relative to spatial coordinates and solar activity parameters:
\begin{equation*}
    \theta_{cs} = \pi/2 - \arctan(\tan \alpha \cdot \sin \phi^{*}) \quad \quad \phi^{*} = \phi + r \frac{\Omega_{\odot}}{u},
\end{equation*}
where the tilt angle $\alpha$ determines the current sheet's tilt relative to the solar equatorial plane at $1$ solar radius. The current sheet structure in the XZ plane is compared in Figure 6 for tilt angles $\alpha = 0^{\circ}$ (flat) and $\alpha = 30^{\circ}$ (wavy).

\begin{figure}[!h]
\centering
\includegraphics[width=1\linewidth]{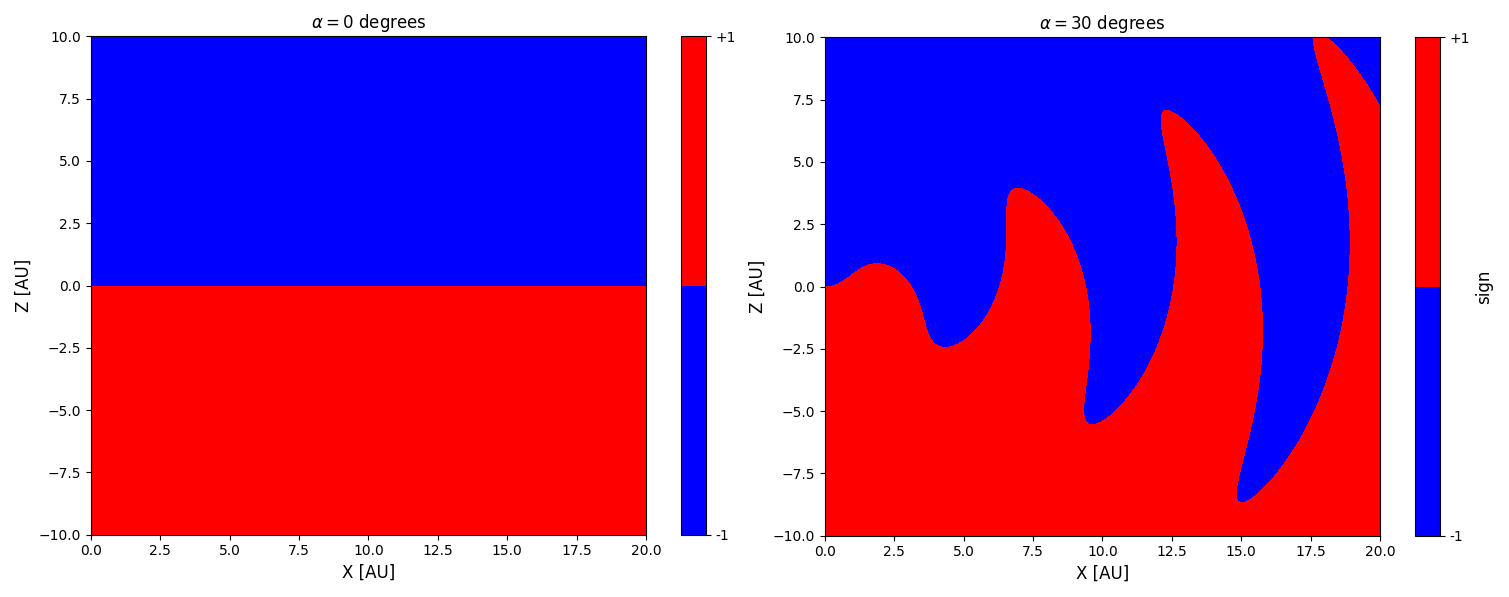}
\caption{Sign of the IMF ($\pm 1$) in the XZ plane (coordinates $(0,0)$ correspond to solar position) during negative polarity ($A_{\text{pol}} = -1$) periods for tilt angles $\alpha = 0^{\circ}$ and $\alpha = 30^{\circ}$.}
\label{fig:mpr}
\end{figure}

The mathematically equivalent system of SDE corresponding to Parker's equation (1) in the general problem formulation takes the form:
\begin{equation}
    d\mathbf{X} = (\nabla \cdot \boldsymbol{\hat{\kappa}} - \mathbf{u} - \mathbf{V_{dr}})ds + \sum_{\sigma} \boldsymbol{\alpha}^{\sigma} dW^{\sigma}(s), \quad \quad  dp = \frac{1}{3}p (\nabla \cdot \mathbf{u}) ds,
\end{equation}
where the vectors $\boldsymbol{\alpha}^{\sigma}$ are defined by the decomposition of the diffusion tensor $\boldsymbol{\hat{\kappa}}$, satisfying $2 \kappa_{ij} = \sum_{\sigma} \alpha_{i}^{\sigma} \alpha_{j}^{\sigma}$. These vectors correspond to the columns of the lower-triangular Cholesky \cite{Cholesky-1924} decomposition matrix for the symmetric, positive-definite matrix $2\boldsymbol{\hat{\kappa}}$. In Cartesian coordinates, the diffusion tensor can be expressed as: 
\begin{equation*}
    \kappa_{ij} = \kappa_{\perp} \delta_{ij} + \frac{(\kappa_{\parallel} - \kappa_{\perp}) B_{i} B_{j}}{B^2}
\end{equation*}
To enable direct comparison with Kota \& Jokipii~\cite{Jokipii-1983}, we adopted identical parameters: outer boundary condition  $f(R,p) \sim (m_p^2c^4 + p^2c^2)^{-1.8}/pc$ at $R = 10\ \text{AU}$ and the absorbing inner boundary condition $f(r_0,p) = 0$ at $r_0 = 0.1\ \text{AU}$, with diffusion coefficients
\begin{equation*}
\kappa_{\parallel} = 50 \times 10^{20} \cdot\frac{v}{c}\left(\frac{p}{p_0}\right)^{1/2}\left(\frac{B_e}{B}\right)\ \text{cm}^2/\text{s}, \quad \kappa_{\perp} = 0.05 \cdot \kappa_{\parallel},
\end{equation*}
where $v$ is particle velocity, $p_0 = 1\ \text{GeV/c}$, and $B_e = \sqrt{2}B_0 \approx 50\ \mu\text{G}$.

Since the characteristic penetration time for a proton with a momentum of $p = 1\ \text{GeV/c}$ from $10\ \text{AU}$ to $1\ \text{AU}$ is on the order of several days, we adopt a steady-state approximation in our model. This implies that the configuration of the heliospheric current sheet remains effectively fixed throughout our computations. Figure 7 demonstrates good agreement between our solutions using the SDE method (solid curves) and the finite-difference results from \cite{Jokipii-1983} and \cite{Burger-2012} (data points), despite differing treatments of drift along the heliospheric current sheet. Notably, these prior studies used a frame co-rotating with the Sun, with results averaged over a solar rotation at 1 AU at the solar equator, whereas our approach adopts a fixed heliocentric frame and performs azimuthal angle averaging at 1 AU at the solar equator.
\begin{figure}[!h]
\centering
\includegraphics[width=0.6\linewidth]{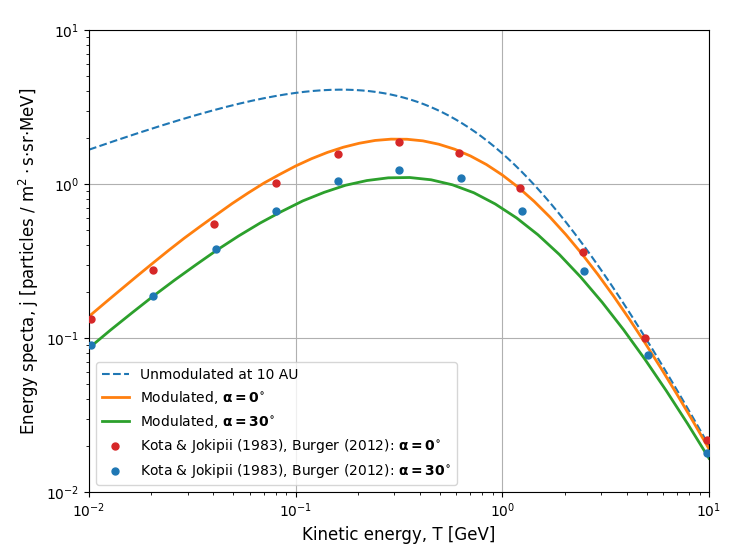}
\caption{Proton energy spectra at $1\ \text{AU}$ at the solar equator (modulated) and $10\ \text{AU}$ (unmodulated) for tilt angles of $0^{\circ}$ and $30^{\circ}$ under negative polarity ($A_{\text{pol}} = -1$). Solid curves: numerical solutions obtained using the SDE method; data points from \cite{Jokipii-1983} and \cite{Burger-2012}.}
\label{fig:mpr}
\end{figure}

\section{CONCLUSIONS}

In this study, we have developed and validated numerical methods for solving the Parker transport equation, employing both finite-difference (Crank-Nicolson scheme) and SDE approaches. For simplified problem formulations, the results obtained from both methods exhibit excellent agreement with the analytical solution, confirming their accuracy. For the more complex case incorporating three-dimensional geometry, anisotropic diffusion, and drift effects, we implemented the SDE method, which offers computational efficiency and flexibility. Our results are consistent with previous studies \cite{Jokipii-1983, Burger-2012}, validating our solution methodology and the correct implementation of drift effects along the heliospheric wavy current sheet.

Future research will focus on applying the SDE method to study GCR modulation through the global heliosphere, including the heliospheric shock structure. These investigations will be conducted using a state-of-the-art global heliospheric model \cite{Izmodenov-2020}.

\section{CONFLICT OF INTEREST}
The author of this work declares that he has no conflicts of interest.

\end{document}